\documentclass[graybox]{svmult}
\usepackage[utf8]{inputenc}

\title{HLRS2021}
\author{xu.chu }

\usepackage{float}
\usepackage{mathptmx}       
\usepackage{helvet}         
\usepackage{courier}        
\usepackage{type1cm}        

\usepackage{makeidx}         
\usepackage{graphicx}        
\usepackage{subfigure}
\usepackage{caption}
\usepackage{multicol}        
\usepackage[bottom]{footmisc}
\usepackage{multirow}
\usepackage{latexsym}
\usepackage{epstopdf}
\usepackage[square]{natbib}
\usepackage{authblk}
\usepackage{amsmath,bm}

\begin{document}

\title*{An investigation of information flux between turbulent boundary layer and porous medium}
\titlerunning{Turbulence modulation and energy transfer above porous media}

\authorrunning{X.Chu, W. Wang,  B.Weigand \email{user@gmail.com}}

\author{Xu Chu \inst{1} \and Wenkang Wang \inst{2} \and Bernhard Weigand \inst{3}}
\institute{\inst{1} Cluster of Excellence SimTech, University of Stuttgart, Pfaffenwaldring 5a, 70569 Stuttgart, Germany,  \email{xu.chu@simtech.uni-stuttgart.de} \and %
                      \inst{2} Max Planck Institute for Intelligent Systems, Heisenbergstraße 3, 70569 Stuttgart, Germany \and %
                    \inst{3} Institute of Aerospace Thermodynamics (ITLR), Pfaffenwaldring 31, 70569 Stuttgart, Germany  
                      }
\authorrunning{X.Chu, W. Wang,  B.Weigand}                      

\maketitle
\vspace{-3cm}
\abstract {
The interaction between boundary layer turbulence and a porous layer is the cornerstone to the interface engineering.  In this study, the spatial resolved transfer entropy is used to assess the asymmetry of the causal interaction next to a permeable wall. The analysis was based on pore-resolved direct numerical simulation of turbulent channel flow over a cylinder array. The spatial map of transfer entropy reveals the information flux between the porous medium and arbitrary nearby position. }

\,

\section{Introduction}\label{sec:1}

Turbulent flow over permeable interfaces is ubiquitous in nature and engineering applications, from sediment transport to the transpiration cooling in gas turbines. Porous media are generally featured with complex topology in space \citep{Chu.2018}. A wide range of properties has been identified to have a significant effect on the mass and momentum transport across the porous medium-free flow interface. However, understanding and optimizing macro/micro-scale characteristics of porous media remains a challenge. This is due to both the difficulty of experimental measurements at the pore scale \cite{Terzis.2019} as well as the high costs of numerical simulations to resolve the full spectrum of scales.


Early studies of turbulent flow over permeable beds have addressed the effect of varying permeability on surface flows. The turbulent surface flow is similar to that of a canonical boundary layer when the
interfacial permeability is low, as the near-wall structures are less disrupted.  However, as the permeability increases, large-scale vortical structures emerge in the surface flow. This is attributed to Kelvin-Helmholtz (KH) type instabilities from inflection points of the mean velocity profile. More recently, the application of porous media for drag reduction inspired a series of in-depth studies on anisotropic permeability. 
Direct Numerical Simulation (DNS) exhibits an edge of observing and analyzing turbulent physics in a confined small space, not only for the canonical cases as channel flows and pipe flows \cite {Chu.2016, Chu.2020b,Mceligot.2018, Mceligot.2020, Pandey.2017b}.
Suga et al. \cite{Suga.2020} investigated the influence of the anisotropic permeability tensor of porous media at a higher permeable regime. It was found that streamwise and spanwise permeabilities enhance turbulence whilst vertical permeability itself does not. In particular, the enhancement of turbulence is remarkable over porous walls with streamwise permeability, as it allows the development of streamwise large-scale perturbations induced by the Kelvin–Helmholtz instability. 

The concept of transfer entropy was proposed \cite{Schreiber2000} as a tool to evaluate the directional information transfer between a source signal and a target signal. This definition of transfer entropy measures the information contained in the source about the next state of the target that was not already contained in the target's past. This allows one to differentiate the direction of the information flux, and can therefore be used to quantify causal interactions. Recently, Lozano-Dur{\'a}n \cite[]{lozano2020causality,lozano2020cause,lozano2022information} highlighted the importance of causal inference in fluid mechanics and proposed leveraging information-theoretic metrics to explore causality in turbulent flows. Wang et al. \cite{Wang.2021} used transfer entropy as a marker to evaluate the causal interaction between turbulent channel flow and porous media consisting of circular cylinders. The POD time coefficients of the leading order modes were extracted as the representative signals for surface and sub-surface flow. 
 
The explorations above illustrate the potential strength of information-theoretic tools in revealing the surface/sub-surface coupling dynamics. However, there are still numerous challenges and unanswered questions. In the current work, we will compute spatial resolved transfer entropy between turbulent fluctuations in channel flows over porous media. 
\section{Numerical method}
\label{sec:2}

\begin{figure}
	\begin{tabular}{cc}
	 \includegraphics[width=0.42\textwidth]{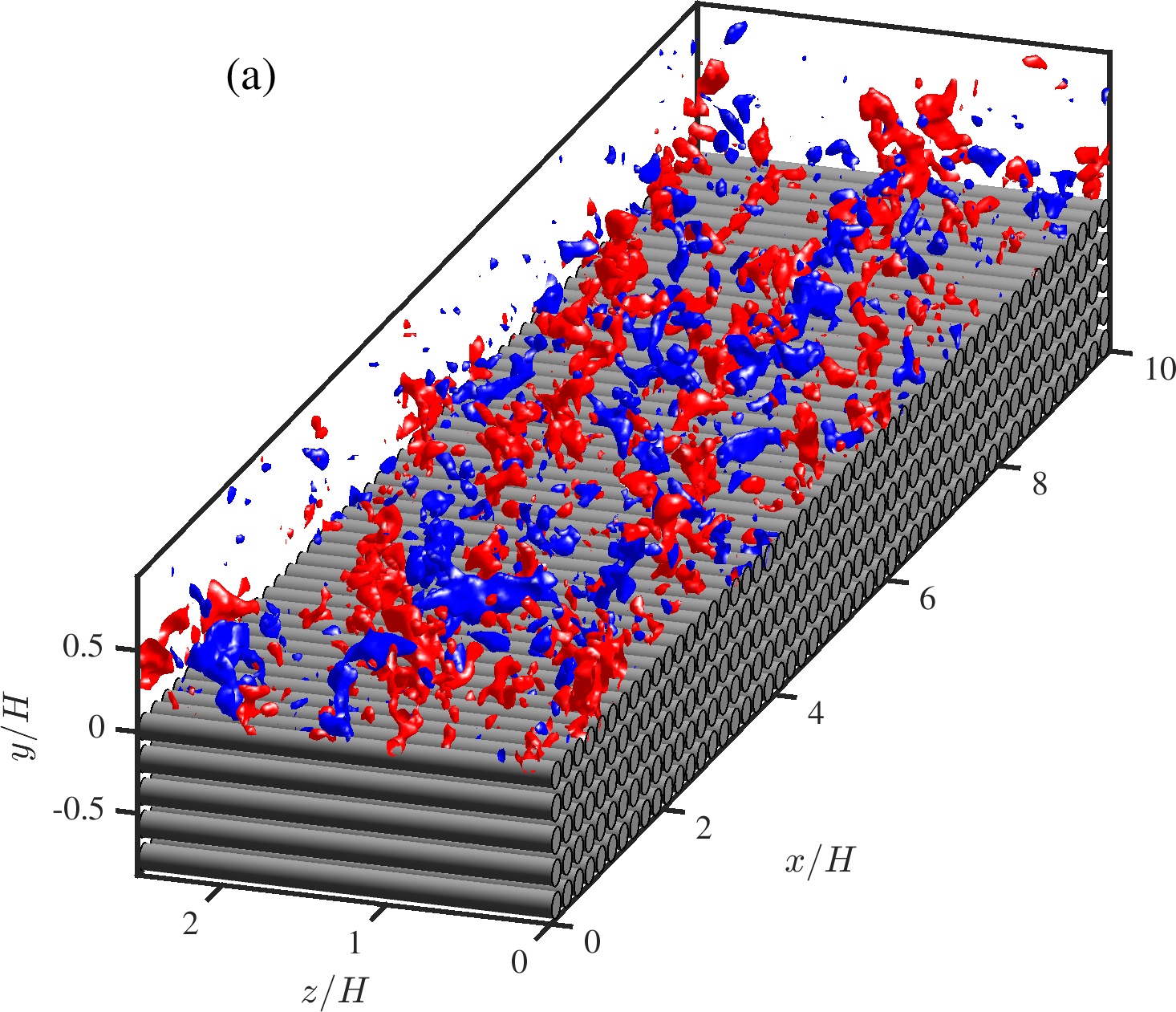}    &  
	 \includegraphics[width=0.55\textwidth]{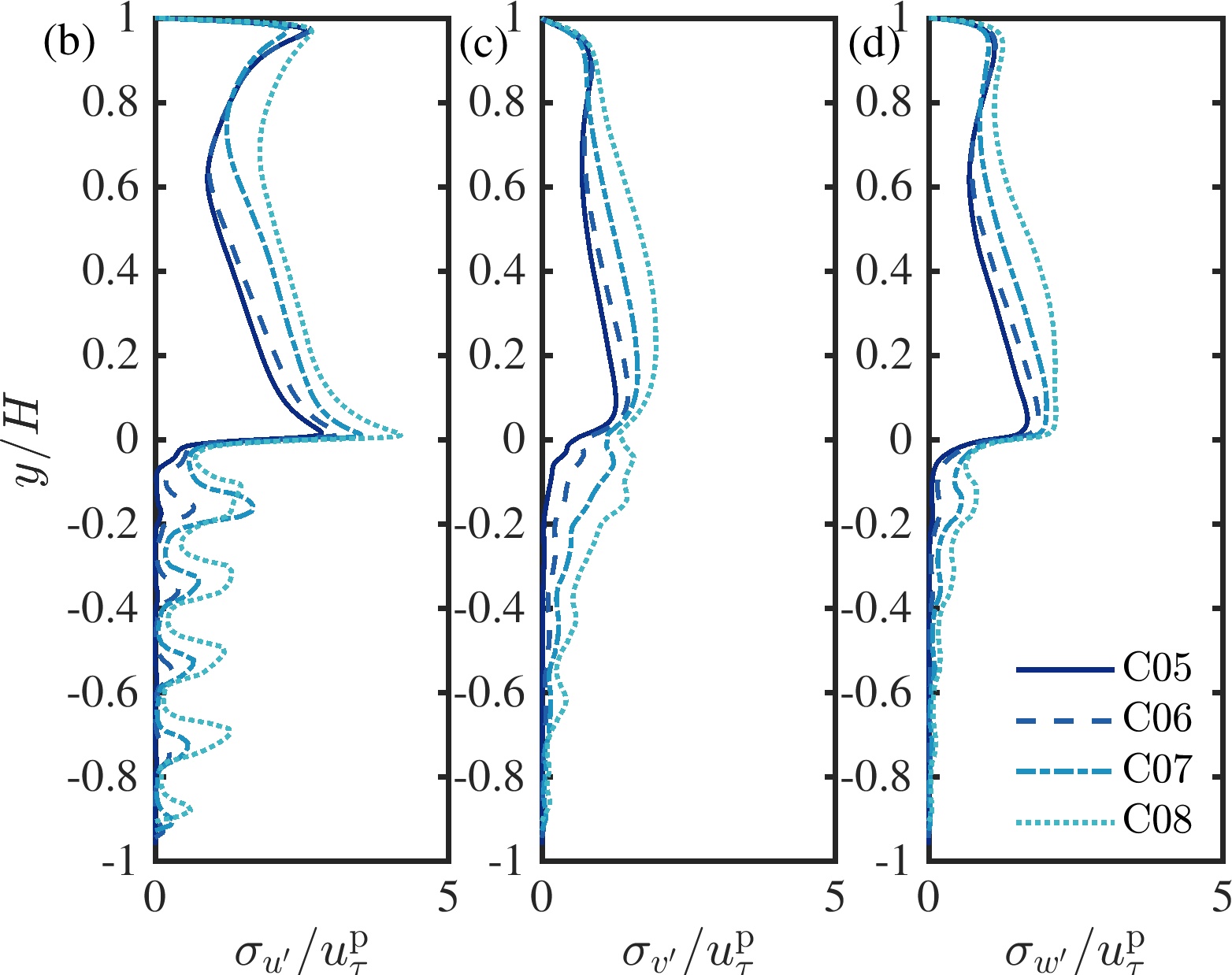}
	\end{tabular}
	
	\caption{(a) Configuration of the computational domain (C05). The blue and red isosurfaces show the wall-normal fluctuation $v'/u^p_\tau$ at level $-0.4$ and $0.4$, respectively. Panels (b-d) show the profiles of streamwise $u'$, wall-normal $v'$, and spanwise $w'$ fluctuation intensity, respectively. }
	\label{fig:setup}
\end{figure}

In our DNS, three-dimensional incompressible Navier$-$Stokes equations are solved in a non-dimensional form without models, 

\begin{equation}
\frac{\partial u_j}{\partial x_j}=0
\label{eq:1}
\end{equation}
\begin{equation}
\frac{\partial u_i}{\partial t}+\frac{\partial u_i u_j}{\partial x_j}=-\frac{\partial p}{\partial x_i}+\frac{1}{Re_D}\frac{\partial^2 u_j} {\partial x_i \partial x_j}+\Pi \delta_{i1}
\label{eq:2}
\end{equation}

where $\Pi$ is a constant pressure gradient in the mean-flow
direction.  The governing equations are non-dimensionalized by normalizing lengths by the half-width of the whole simulation domain $H$(figure \ref{fig:setup}a), velocities by the averaged bulk velocity $U_b$ of the free flow region ($y/H=[0,1]$), such that time is non-dimensionalized by $H/U_b$.
The spectral/$hp$ element solver Nektar++
\citep{Cantwell.2011, Chu.2019, Chu.2020, Chu2021transport, chu2021turbulence, Pandey.2020, wang2021b} is used to discretize the numerical domain
containing complex geometrical structures. The solver allows for arbitrary-order spectral/$hp$ discretizations with hybrid-shaped elements.  The time-stepping is performed with a second-order mixed implicit-explicit (IMEX) scheme. The time step is fixed to $\Delta t/(H/U_b)=5\times10^{-4}$.
  

Hereafter, the velocity components in the streamwise $x$, wall-normal
$y$, and spanwise $z$ directions are denoted as $u$, $v$, and $w$,
respectively.  The domain size ($L_x/H \times L_y/H \times L_z/H$) is $10
\times 2 \times 0.8\pi$ in all cases. The lower half ($y/H=[-1,0]$) contains the porous
media and the upper half ($y/H=[0,1]$) is the free channel flow. The
  porous layer consists of 50 cylinder elements along the
  streamwise direction and 5 rows in the wall-normal
  direction as illustrated in figure \ref{fig:setup}. The distance
  between two nearby cylinders is fixed at $D/H=0.2$. No-slip boundary conditions are applied to the cylinders, the upper wall, and the lower wall. Periodic boundary conditions are used in
streamwise and spanwise directions. The geometry is discretized using
quadrilateral elements on the $x-y$ plane with local refinement near
the interface. Lagrange polynomials (polynomial order $P=5-9$) are used within each element on the $x-y$ plane. The spanwise direction is extended with a Fourier spectral method.

\begin{table}
\centering
\begin{tabular}{c|cccccccccc}
\hline
case&$\varphi$  & $Re_{\tau}^{\mathrm{p}}$&$\Delta x^{\mathrm{p}+}$/$\Delta y^{\mathrm{p}+}$/$\Delta z^{\mathrm{p}+}$&$Re_{\tau}^{\mathrm{s}}$&$\Delta x^{\mathrm{s}+}$/$\Delta y^{\mathrm{s}+}$/$\Delta z^{\mathrm{s}+}$&$\sqrt{K_{xx}}^{\mathrm{p}+}$,$\sqrt{K_{yy}}^{\mathrm{p}+}$ 
 &$\sqrt{K_{zz}}^{\mathrm{p}+}$ & $C_{xx}^{\mathrm{p+}},C_{yy}^{\mathrm{p+}}$ & $r^{\mathrm{p+}}$\\
\hline
C05  &0.5 & 336 & 2.2/0.38/5.3 & 180 & 6.3/0.43/4.5 &4.55&8.86&0.37&42\\
C06  & 0.6 & 464 & 3.5/0.50/6.8  & 190 & 5.5/0.47/5.4 &9.34&15.23&0.58&48 \\
C07   &0.7   & 625 & 4.2/0.55/8.4  & 160 & 5.2/0.51/5.1 &20.65&30.98&2.99&52\\
C08   &0.8   & 793 & 3.1/0.51/5.0  & 170 & 4.1/0.40/2.8 &36.83&53.99&13.58&52\\
\hline
\end{tabular}
\caption{Simulation parameters. The porosity of the porous medium
  region is $\varphi$. The friction Reynolds numbers are $Re_{\tau}^{\mathrm{p}}$ and
  $Re_{\tau}^{\mathrm{s}}$ for the porous and impermeable top walls respectively. $\Delta
  x^{\mathrm{p}+}$/$\Delta y^{\mathrm{p}+}$/$\Delta z^{\mathrm{p}+}$
  and $\Delta x^{\mathrm{s}+}$/$\Delta y^{\mathrm{s}+}$/$\Delta
  z^{\mathrm{s}+}$ are the respective grid spacings in wall units of
  porous wall and top wall, respectively. $\sqrt{K_{\alpha\alpha}}^{\mathrm{p}+}$ and $C_{\alpha\alpha}^{\mathrm{p+}}$ are the diagonal components of the permeability tensor and Forchheimer coefficient, respectively, in the direction of $\alpha$ ($\alpha\in\{x, y,z \}$). $r^{\mathrm{p+}}$ is the radius of the cylinders. All the variables are normalized using $u_\tau^{\mathrm{p}}$ and $\nu$.}
\label{tab:1}
\end{table}

Four DNS cases are investigated with varying porosity $\varphi=0.5,0.6,0.7,0.8$, which is
defined as the ratio of the void volume to the total volume of the
porous structure. The parameters of the simulated cases are listed in table \ref{tab:1}, where the cases are named after their respective porosity. The superscripts $(\cdot)^{\mathrm{p}}$ and
$(\cdot)^{\mathrm{s}}$ represent variables of the permeable wall and the top smooth wall sides, respectively. Variables with superscript $^+$ are scaled by the friction velocities $u_\tau$ of their respective side and the viscosity $\nu$. Note that the distance between cylinders is fixed, and the porosity is changed by varying the radius of the cylinders. 
The normalized cylinder radius is in the range $r^{\mathrm{p}+}=42$--$52$ for all the cases tested (see table \ref{tab:1}) such that the effect of surface roughness is assumed to be on a similar level.  The Reynolds number of the top wall boundary layer is set to be $Re_\tau^{\mathrm s}=\delta^{\mathrm s}u_\tau^{\mathrm
  s}/\nu\approx180$ for all cases ($\delta$ is the distance between the position of maximum streamwise velocity and the wall). 
In this manner, changes in the top wall
boundary layer are minimized. 
  On the top smooth wall side, the streamwise cell size ranges from $4.1\le\Delta
x^{\mathrm{s}+}\le6.3$ and the spanwise cell size is below $\Delta
z^{\mathrm{s}+}=5.4$. On the porous
media side, $\Delta z^{\mathrm{p}+}$ is below 8.4, whereas $\Delta x^{\mathrm{p}+}$ and $\Delta y^{\mathrm{p}+}$ are enhanced by polynomial refinement of a local mesh \citep{Cantwell.2011}. The total number of grid points ranges from 88$\times10^6$ (C04) to 595$\times10^6$ (C08). Each cylinder in the porous domain is resolved with 80 to 120 grids along the perimeter. By performing parameter tests on porous media and fitting the Darcy-Forchheimer equation, the permeability tensor $K$ and Forchheimer coefficient $C$ may be obtained.

The velocity field $\bm{u}(\bm{x},t)=(u,v,w)$ can be further decomposed into the time-averaged velocity $\overline{\bm{u}}=\int {\bm{u}}\mathrm{d}t/T$ and turbulent fluctuation $\bm{u}^\prime=(u^\prime,v^\prime,w^\prime)=\bm{u}-\overline{\bm{u}}$. Figure \ref{fig:setup}(a) shows a snapshot of wall normal fluctuation $v\prime$. The intensity profiles of $u^\prime$, $v^\prime$, and $w^\prime$ are shown in Fig.\ref{fig:setup}(b-d), respectively. The turbulent intensity grows with porosity for both the channel and porous medium region. In the current study, we focus on the up and down-welling motions at the upper surface of the porous medium, which is represented by the fluctuation $v'$. 

\section{Results and analysis}
\label{sec:3}

\subsection{Transfer entropy}

In the current work, the transfer entropy \citep{Schreiber2000, lozano2020cause, lozano2022information, williams2022information, arranz2022assessment} is used to evaluate the direction of coupling, i.e., the cause-effect relationship, between two time-series. The transfer entropy originates from the framework of information theory, and is a specific version of the mutual information for conditional probability. It represents the dependence of the future state of process $X$ on the process $Y$, which is measured as the decrease of uncertainty (entropy) of $X$ by knowing the past state of $Y$. The transfer entropy from $X$ and $Y$ can be defined as:

\begin{equation}
    T_{Y\rightarrow X}(\Delta t)=H(X_t|X_{t-1})-H(X_t|{X_{t-1},Y_{t-\Delta t}})
    \label{eq:Tij}
\end{equation}
where $\Delta t$ is the time-lag to evaluate causality, and $H(A|B)$ is the conditional Shannon entropy of the variable $A$ given $B$, which is defined as 

\begin{equation}
    H(A|B)=-E\left[\mathrm{log}(p(A|B)\right]=-E\left[\mathrm{log}\left(\frac{p(A,B)}{p(B)}\right)\right]=E[\mathrm{log}_2(p(B))]-E[\mathrm{log}_2(p(A,B))]
    \label{eq:H}
\end{equation}
where $p(\cdot)$ is the probability density function, and $E[\cdot]$ denotes the expectation value. $H(A|B)$ denotes the chunk of information in $A$ that has nothing to do with $B$. 

In order to quantify the strength of causality, normalization is necessary to scale the magnitude of causality within $[0,1]$ and eliminate small values caused by statistical errors. The normalized transfer entropy can be defined as:

\begin{equation}
    \tilde T_{Y\rightarrow X}=\frac{T_{Y\rightarrow X}-E[T_{Y^s\rightarrow X}]}{H(X_t|X_{t-1})}
    \label{eq:nte}
\end{equation}
where $E[T_{Y^s\rightarrow X}]$ is an estimation of the statistical bias. $Y^s$ is the surrogate variable of $Y$, which is acquired by randomly permuting $Y$ in time to break its causal links with $X$. The conditional entropy $H(X_t|X_{t-1})$ is the intrinsic uncertainty of $X$ knowing its history. Eq.\ref{eq:nte} thus represents the fraction of information in the target $X$ not explained by its past that is explained by $Y$ in conjunction with that past.

\subsubsection{Local transfer entropy}

The physical interpretation of the quantity of transfer entropy can be hard to grasp and sometimes the outcome is confused with the more-widely used correlation method. In fact, the two methods are indeed closely related. In the following, the results from both analysing tools will be presented and compared. In the current section, we will present a simple example to illustrate the strength of transfer entropy and explore the connection between the information transfer and the flow motion.

The definition of transfer entropy quantifies the statistical coherence between time-evolving systems in a global and averaged manner. Using Eq.\ref{eq:H}, we rewrite Eq.\ref{eq:Tij} in the following form:

\begin{equation}
    T_{Y\rightarrow X}(\Delta t)=-E\left[\frac{\mathrm{log}_2(p(X_t|X_{t-1}))}{\mathrm{log}_2(p(X_t|X_{t-1},y_{t-\Delta t}))}\right]
    \label{eq:Tij2}
\end{equation}

We define the local transfer entropy $\zeta$ as 

\begin{equation}
\zeta_{Y \rightarrow X}(t,\Delta t)=\log _{2} \frac{p\left(x_{t} \mid x_{t-1}, y_{t-\Delta t}\right)}{p\left(x_{t} \mid x_{t-1}\right)}
\label{eqn:lte}
\end{equation}
where $x_{t}$ and $y_{t}$ are instantaneous measurements at time $t$ of processes $X$ and $Y$, respectively. The transfer entropy in Eq.\ref{eq:Tij} is essentially the expectation value of the local transfer entropy, that is:

\begin{equation}
T_{Y \rightarrow X}=E[ \zeta_{Y \rightarrow X}(t, \Delta t)]
\end{equation}

\begin{figure}
\centering
\includegraphics[width=0.7\textwidth]{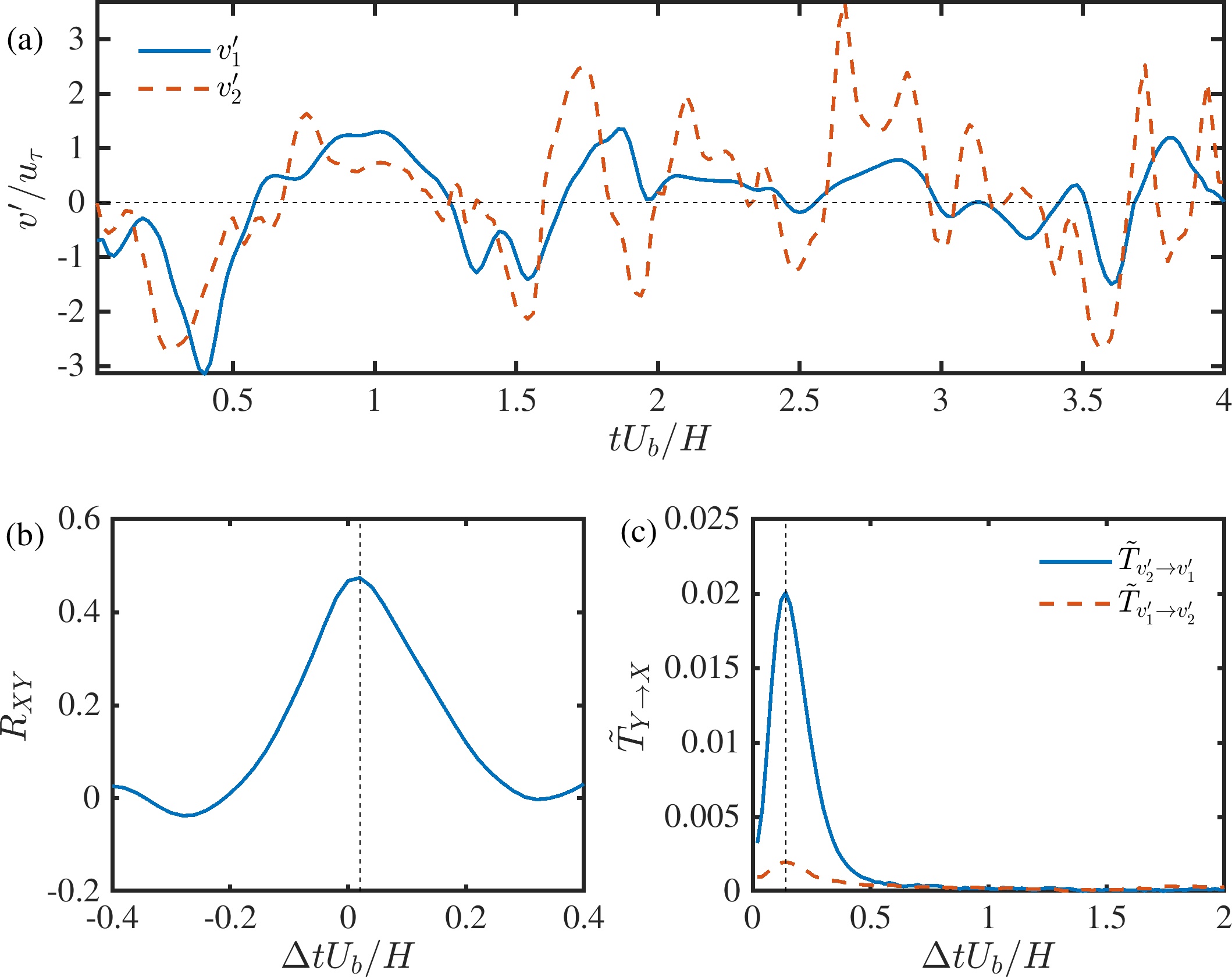}\\
\caption{Comparison between correlation and transfer entropy. (a) Excerpts of instantaneous vertical fluctuations $v_1'$ and $v_2'$ at $y=-r_c$ (solid line) and $y=0$, respectively; (b) profile of cross-correlation between $v_1'$ and $v_2'$. (c) Transfer entropy $\tilde T_{v_2'\rightarrow v_1'}$ (solid line) and $\tilde T_{v_1'\rightarrow v_2'}$ (dashed line) as a function of time delay $\Delta t$.}
\label{fig:local1}
\end{figure}

To explain the idea of (local) transfer entropy more explicitly,  figure \ref{fig:local1}(a) shows the excerpts of wall-normal fluctuation $v_1'(t)$ and $v_2'(t)$ of case C05, which are extracted from the middle of the first-layer cylinder gaps ($y=-r_c$) and crest position ($y=0$), respectively. The two signals $v_1'$ and $v_2'$ are acquired from the same streamwise position. Despite their clear differences in scale content, the two signals appear to be connected, with the positive and negative peaks appearing to be coordinated. This feature is also evidenced by the cross-correlation profile between $v_1'$ and $v_2'$ (see Fig.\ref{fig:local1}b), which is defined as

\begin{equation}
  R_{vv}(\Delta t)=\overline{v_1'(t)v_2'(t+\Delta t)}/(\sigma_{v_1'}\sigma_{v_2'}),
\end{equation}
where $\Delta t$ is the shifted time interval between the two signals. The profile shows a positive peak value of 0.48 at $\Delta tU_b/H\approx0.02$, revealing that the two signals are largely correlated. However, a high correlation value does not always reflect information flow or a strong causal relationship because $R$ contains no discrimination about direction. As a comparison, figure \ref{fig:local1}(c) shows the transfer entropy $\tilde T_{v_2'\rightarrow v_1'}$ and $\tilde T_{v_1'\rightarrow v_2'}$ as functions of $\Delta t$. First of all the time delay for transfer entropy can only be positive due to the constraint that the `cause' always happens before the `effect'. Secondly, there is a huge disparity between the magnitude of $\tilde T_{v_2'\rightarrow v_1'}$ and $\tilde T_{v_1'\rightarrow v_2'}$, which reveals the strong dependence of information flux on the coupling direction. For the current signal set, the transfer entropy in both directions reaches a maximum at $\Delta tU_b/H\approx 0.15$.  

\begin{figure}
\centering
\includegraphics[width=0.75\textwidth]{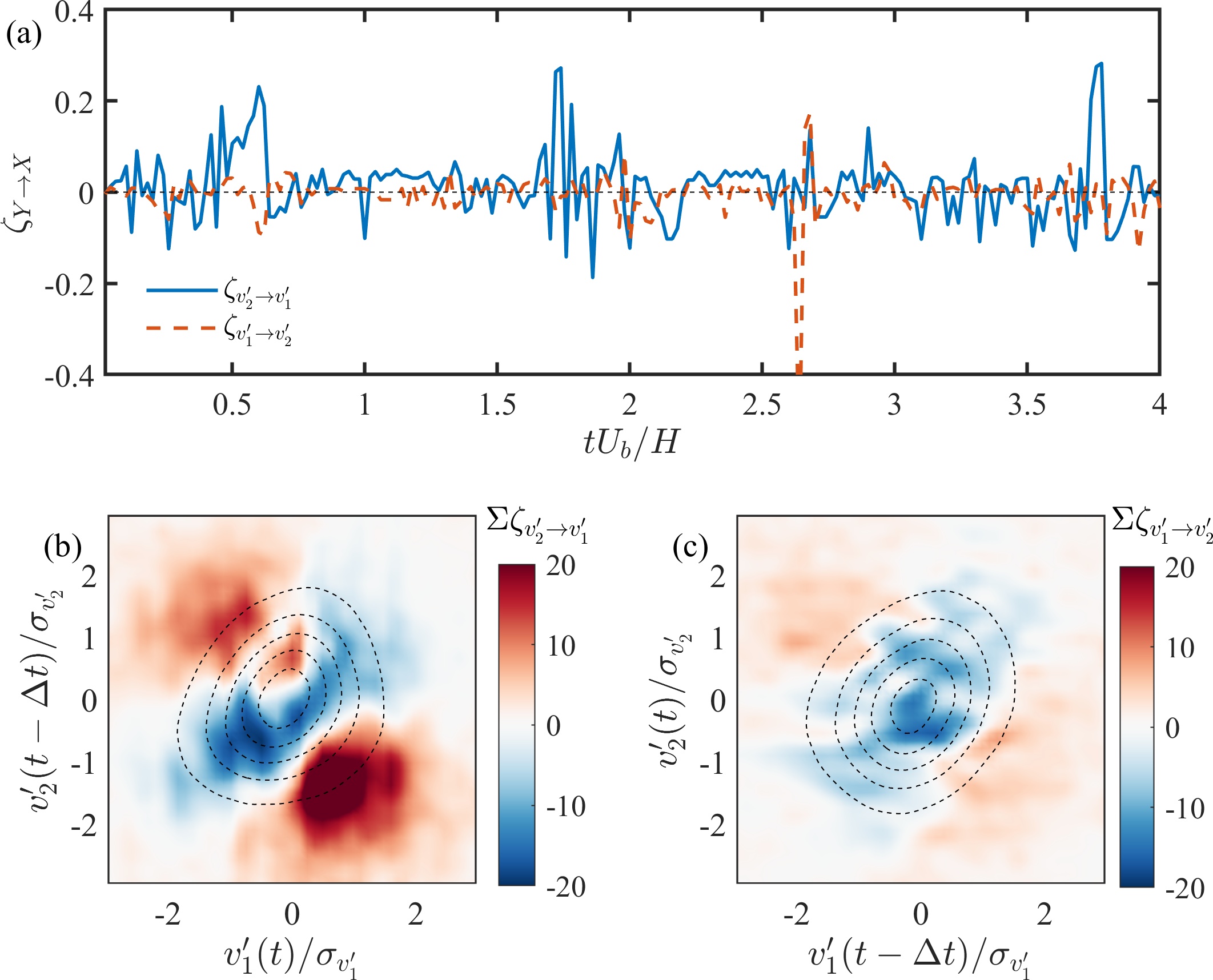}
\caption{Interpretation of transfer entropy from the scope of local events. (a) local transfer entropy $\zeta_{v_2'\rightarrow v_1'}$ (solid line) and $\zeta_{v_1'\rightarrow v_2'}$ (dashed line) corresponding to the time series shown in Fig.\ref{fig:local1}(a). The colour in panels (b) and (d) show the accumulative maps of $\zeta_{v_2'\rightarrow v_1'}$ and $\zeta_{v_1'\rightarrow v_2'}$ in the space of $(v_1',v_2')$, respectively. The isolines are the joint probability density function of $p(v_1',v_2')$ with a time delay $\Delta t$ imposed on the source signal. The levels are from 1$\times10^{-3}$ to 3$\times10^{-3}$ with a step of 5$\times10^{-4}$.}
\label{fig:local2}
\end{figure}

To further explain the concept of transfer entropy, Fig. \ref{fig:local2}(a) shows the local transfer entropy corresponding to the time-series in Fig.\ref{fig:local1}(a). The time delay $\Delta t$ is fixed at $\Delta tU_b/H=0.15$, corresponding to the maximum value of transfer entropy. There are several things to be noted. First, the $\zeta$ of the two transfer directions are totally different. The magnitude of $\zeta_{v_2'\rightarrow v_1'}$ is generally larger than that of $\zeta_{v_1'\rightarrow v_2'}$, which explains the disparity between the two profiles of $\tilde T$ in Fig.\ref{fig:local1}(c). Secondly, both position and negative values exist for local transfer entropy. Unlike the global transfer entropy that is positive by definition \cite[]{Schreiber2000}, the value of local transfer entropy is not
constrained. A positive $\zeta(t,\Delta t)$ indicate an informative source for a speciﬁc event set $\{x_t, x_{t-1}, y_{t-\Delta t}\}$, while a negative $\zeta(t,\Delta t)$ occurs when the probability of observing the actual next state of the destination given the value of the source $p\left(x_{t} \mid x_{t-1}, y_{t-\Delta t}\right)$, is lower than the probability knowing only the history of the destination $p\left(x_{t} \mid x_{t-1}\right)$. In this case, the source element is actually misleading regarding the destination's state transition.

In the brief excerpt of $\zeta_{v_2'\rightarrow v_1'}$ (Fig. \ref{fig:local2}a), we observed several major positive peaks, when both $v_1'$ and $v_2'$ are of strong magnitude. However, this can not be taken as an evidence that positive local transfer entropy captures the shifted peaks between $v_1'$ and $v_2'$, as there are also locations where shifted peaks do not generate strong positive $\zeta$. To understand the true event contributing to positive $\zeta$, we computed the accumulative map of $\zeta$ in the space of $(v_1',v_2')$ (Fig.\ref{fig:local2}b,c). In addition, the joint probability density function (JPDF) $p(v_1',v_2')$ is superimposed on the accumulative map as dashed isolines for comparison.

It is shown in Fig.\ref{fig:local2}(b) that $\zeta_{v_2'\rightarrow v_1'}$ is mainly positive in the second (i.e., $(v_1'(t)<0,v_2'(t-\Delta t)>0)$) and fourth quadrants (i.e., $(v_1'(t)>0,v_2'(t-\Delta t)<0)$), and negative in the remaining areas. Moreover, the events with $v_1'$ and $v_2'$ of a small magnitude are mostly `misleading', although they are of a larger number (see the JPDF indicated by dashed isolines). We focus on the major positive peak in the 4th quadrant as it represents dominant scenario of the predictive event sets. The peak locates at $(v_1'(t)/\sigma_{v_1'},v_2'(t-\Delta t)/\sigma_{v_2'})=(0.5,-1.5)$. This is consistent with the positive $\zeta_{v_2'\rightarrow v_1'}$ peaks (indicated by arrows) in Fig.\ref{fig:local1}(a), where a large negative $v_2'$ happens first, and then a minor positive $v_1'$ is found after a time delay of around $\Delta tU_b/H=0.15$. Such a combination of source and target events is essentially different from the events contributing to the correlation peaks in Fig.\ref{fig:local1}(b), i.e., those slightly shifted peaks. During the computation of transfer entropy, the information carried by the immediate past of the destination is subtracted from the predictive information. This may explain why the shifted peaks are not identified as informative events by transfer entropy, as the information of about the peak could be already encoded in the history of the destination signal. The accumulative map of $\zeta_{v_1'\rightarrow v_2'}$ (Fig.\ref{fig:local2}c) shares a similar pattern with that of $\zeta_{v_2'\rightarrow v_1'}$ in Fig.\ref{fig:local2}(b), but with a much smaller magnitude. This is consistent with the previous observation of Fig.\ref{fig:local2}(a).

\subsection{Inter-layer correlation and transfer entropy}\label{sec:corr}

\begin{figure}
	\centering
	\includegraphics[width=0.7\textwidth]{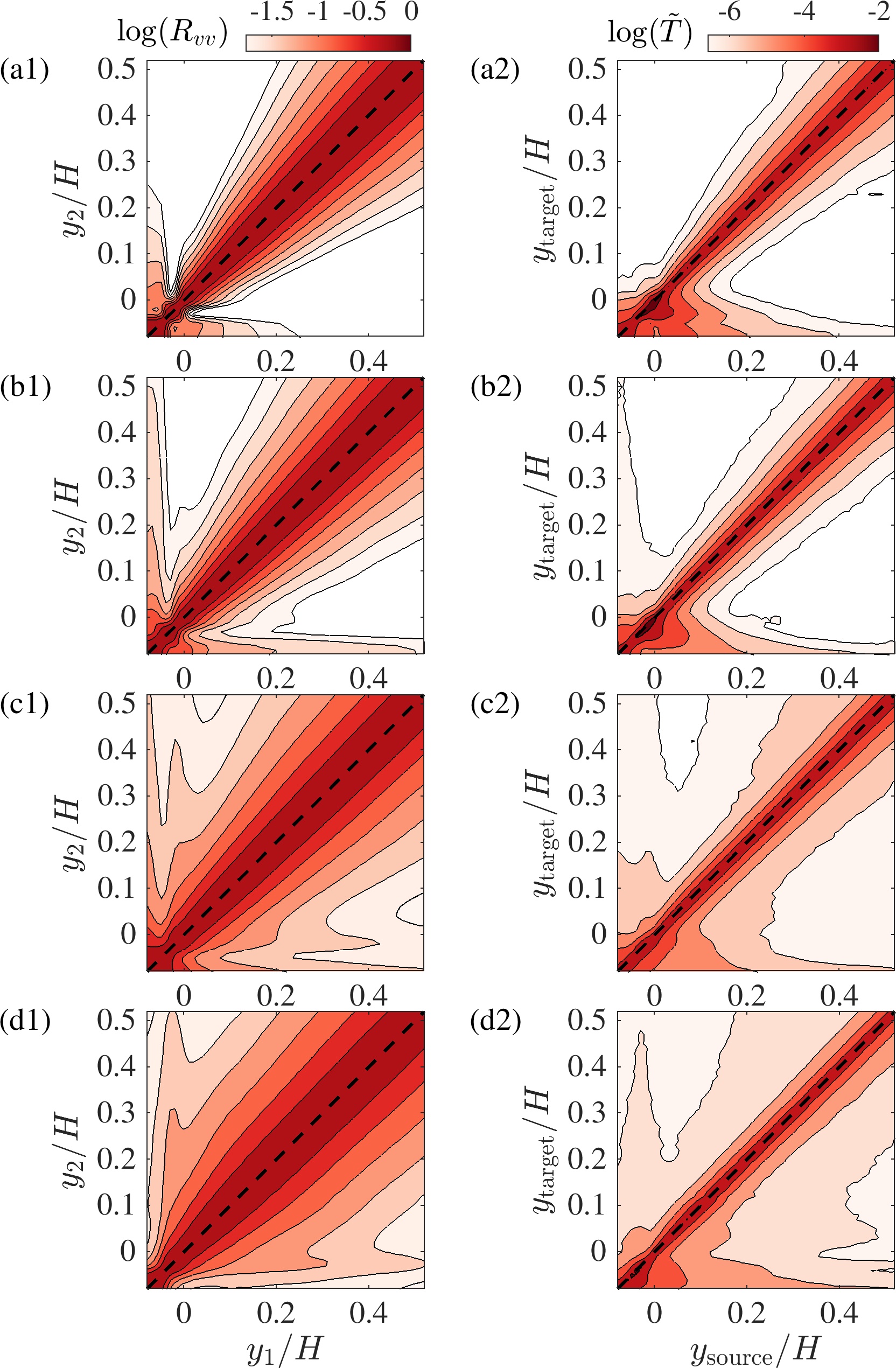}
	\caption{ (a1-d1) Map of inter-layer correlation coefficient of $v'$ fluctuations. Panels (a1-d1) represent cases C05-C08, respectively. The color and isolines show the logarithm of the correlation coefficient, with levels from -1.5 to -0.01 with a step of 0.02. (a2-d2) Map of transfer entropy between different $y$ positions. Panels (a2-d2) represent cases C05-C08, respectively. The color and isolines show the logarithm of the transfer entropy, with levels from -6 to -2 with a step of -1. }
	\label{fig:corr}
\end{figure}

\subsubsection{Correlation map}
To illustrate the statistical connection between wall-normal positions, we calculate the map of maximum correlation coefficient for the $v$ fluctuations at different $y$ positions, which is defined as

\begin{equation}
    R^{\mathrm{max}}_{vv}(y_1,y_2)=\mathop{\mathrm{max}}\limits_{\Delta t}\left\{\frac{\overline{v'(t;y_2)v'(t+\Delta t;y_1)}}{\sigma_{v'}(y_1)\sigma_{v'}(y_2)}\right\}
    \label{eq:R1}
\end{equation}
where $\mathop{\mathrm{max}}\limits_{\Delta t}$ denotes the maximum regarding the variable of $\Delta t$. Figure \ref{fig:corr}(a1-d1) show the contours of $R^{\mathrm{max}}_{vv}(y_1,y_2)$ for all the cases. By definition, the correlation maps are exactly symmetrical about the diagonal line $y_1=y_2$, since the correlation doesn't differentiate the order of the two input signals.   

For lower porosity cases C05 and C06 (Figure \ref{fig:corr}a1,b1), the correlation map shows a `neck' around $(y_1,y_2)=(0,0)$ which separates the contour into two part, i.e., the part with both $y_1$ and $y_2$ above the interface, and the one involving porous medium flow ($y_1<0$ or $y_2<0$). Without losing generality, we can consider $y_2$ (the vertical axis) as the position of the reference $v'$, and its correlated range is shown on $y_1$ axis (the horizontal axis). The fluctuation $v'$ at the interface ($y_2=0$) has a quite strong connection with the flow below the interface ($y_1<0$), but a limited correlated range in the channel region ($y_1>0$). This is probably due to the scale separation between eddies of the near-wall region and the outer layers. As $y_2$ increases ($y_2\ge0$), the correlated range of $y_1$ increases linearly with $y_2$, which is associated with the growth of the scale of coherent structures. The correlated $y_1$ range increases more drastically when $y_2$ descends into the porous media area ($y_2<0$). In fact, the correlation coefficient between surface/subsurface flow can even exceed that of two points in the channel. This is a direct evidence that the motions in the transitional layer of the porous medium are interacting with the flow structures in the boundary layer. As porosity grows (C07 and C08), the intensity of the correlation map increases, and the `neck' between the channel region and the porous media region becomes ambiguous. However, the outline of the correlation map remains similar. 

The correlation map provides a point of view into the connection of signals at different layers. However, as explained earlier, this method can not be used to infer the causal relationship. In the next section, the transfer entropy is used to clarify this problem.

\subsubsection{Transfer entropy map}\label{sec:te_map}

In this section, the same set of data as in Sec.\ref{sec:corr} will be used to compute transfer entropy, and the result is shown in Figure \ref{fig:corr}(a2-d2). Since the input signals for transfer entropy are assigned as target or source, the $y$ positions of the target and source $v'$ signals are defined as $y_{\mathrm{target}}$ and $y_{\mathrm{source}}$, respectively. For each set of $(y_{\mathrm{target}},y_{\mathrm{source}})$, the delay time $\Delta t$ is selected
so that the maximum value of transfer entropy is reached.

In contrast to the correlation map, the transfer entropy map is clearly asymmetrical, especially near and below the crest position. When the destination $y_{\mathrm{target}}$ is located in the channel region, the transfer entropy decays quickly as $y_{\mathrm{target}}$ moves away from $y_{\mathrm{target}}$. In contrast, for a target signal below the interface ($y_{\mathrm{target}}<0$), the information flux from the channel flow decays much slower as the rising of $y_{\mathrm{source}}$, suggesting that the fluctuations below the interface is actively subject to the influence of channel flow. The magnitude of the transfer entropy from porous medium to channel flow is almost negligible for the lowest porosity case C05. As the porosity increases, the information flux from below the interface to above (bottom-up effect) rises accordingly, but the top-down coupling always dominates for the cases tested.  


\subsection{Spatial-resolved transfer entropy}\label{sec:R}
In the previous section, we explored the transfer entropy in $y$ direction, which elucidated the causal relationship between the channel and porous medium flow. Note that only one streamwise location is selected which is right at the middle of two nearby cylinders. In this section, the influence of the streamwise position on the transfer entropy will also be investigated. However, instead of enumerating all possible combinations of spatial locations in sec\ref{sec:te_map}, we specify $v'$ at the throat between the top-layer cylinders ($y=-r_c$) as the representative signal of the porous medium, and explore the information flux between it and nearby spatial points.

\begin{figure}
	\centering
	\includegraphics[width=0.89\textwidth]{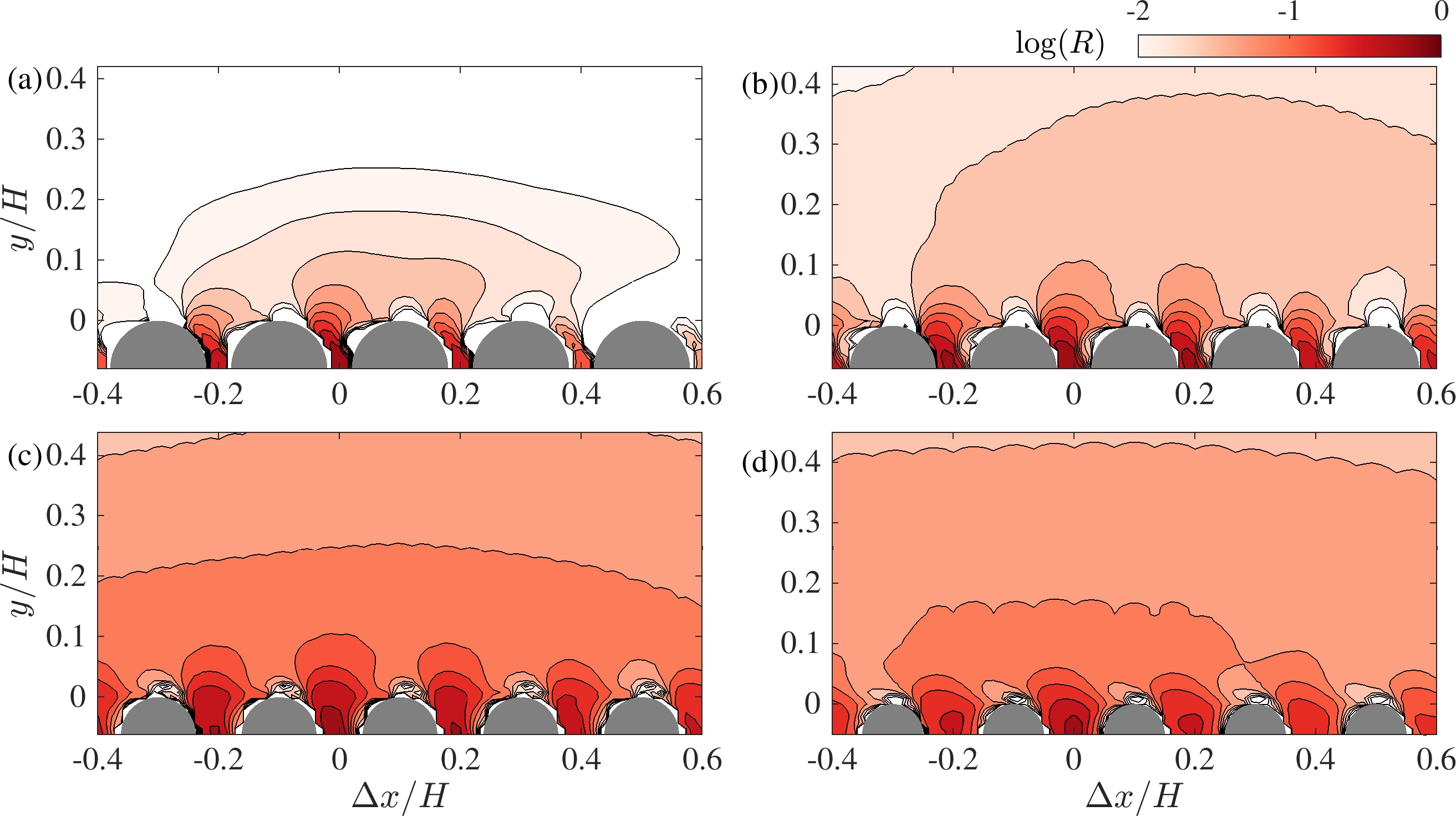}
	\caption{ Map of correlation coefficient $R_{vv}(\Delta x,y)$ in $x-y$ plane. Panels (a-d) represent cases C05-C08, respectively. The reference point is fixed at the middle of cylinders, i.e., $(\Delta x,y)=(0,-r_c)$. The color and the isolines show the logarithm of the correlation coefficient, with levels from -2 to -0.02 with a step of 0.02.}
	\label{fig:R}
\end{figure}

As in the previous sections, we first show the two dimensional correlation map as a comparison with the transfer entropy. With the reference location selected at $(x_{\mathrm{ref}},y_{\mathrm{ref}})$=($x_c,-r_c$), the definition of the correlation map is a little bit different from that in Eq.\ref{eq:R1}, i.e.,

\begin{equation}
        R_{vv}(\Delta x,y)=\mathop{\mathrm{max}}\limits_{\Delta t}\left\{\frac{\overline{v'(t;x_{\mathrm{ref}},y_{\mathrm{ref}})v'(t+\Delta t;x_{\mathrm{ref}},y)}}{\sigma_{v'}(x_{\mathrm{ref}},y_{\mathrm{ref}})\sigma_{v'}(x_{\mathrm{ref}}+\Delta x,y)}\right\}
\end{equation}

The results are shown in Fig.\ref{fig:R}. There is a strong correlation between the $v'$ of nearby upstream and downstream pore units for all cases tested. As discussed by \cite{Wang.2021}, this strong correlation can be attributed to the interaction of fluids in adjacent pores under the constrain of continuity. A strong up-welling motion at one pore could induce down-welling motions of nearby pore units. In addition, Kim et al. \cite{kim2020experimental} show the up-welling and down-welling motions at the permeable surface is under the modulation of large scale structures from the channel flow. Particularly, up-welling motions happen during the passage of a low-speed structure, while down-welling motions usually correspond to high-speed motions. This is also supported by the large scale statistical structure in the channel region, which encapsulate the correlated pore unit below (Fig.\ref{fig:R}).

\section{Computational performance}
\label{sec:5}

The supercomputing systems used for the DNS was {\it HAWK} located at the High Performance Computer Center Stuttgart (HLRS). The new flagship machine {\it HAWK}, based on the Hewlett Packard Enterprise platform running AMD EPYC 7742 processor code named {\it Rome}, has a theoretical peak performance of 26 petaFLOPs, and consists of a 5,632-node cluster. One AMD EPYC 7742 CPU consists of 64 cores, which leads to 128 cores on a single node sharing 256 GB memory on board. This means that a total of 720,896 cores are available on the HAWK system.

\begin{figure}
\centering
\includegraphics[width=4in]{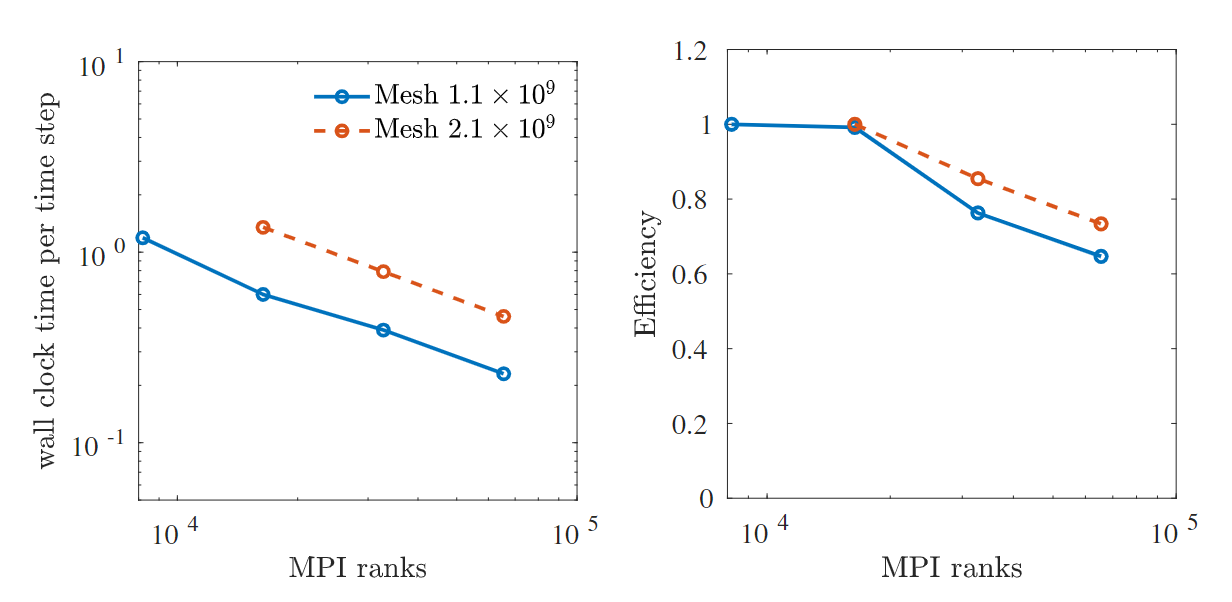}
\caption{Scaling behavior of the solver on ‘HAWK’, wall clock time per time step on the left side. On the right side the efficiency for different numbers of MPI ranks.}
\label{fig:9}
\end{figure}

The scalability of the high-order solver was tested on {\it HAWK} from 8,192 to 65,536 CPU cores. With sufficient computational resource, the wall clock time per time step could reach 0.2-0.4 second, which is optimal to perform direct numerical simulations for turbulent flows. A minimum efficiency of $70\%$ can be achieved with the maximal number of MPI ranks. Compared to the scalability test of the low-order FVM code OpenFOAM \cite{Evrim.2020, Chu.2016, Chu.2016c, Pandey.2017, Pandey.2018, Pandey.2018b, Yang.2020} on the previous HPC machines in HLRS, the current high-order solver shows an excellent scalability on the state of art HPC cluster {\it HAWK}. 

\section{Conclusions}
\label{sec:6}

The causal relationship between surface and sub-surface flow is investigated by using interface-resolved DNS data. Using turbulent fluctuations as the source and target signal, we calculated the transfer entropy between the top-layer pores and a wide range of positions in the channel. The maps of the spatial distribution of transfer entropy illustrate that the information flux from the porous medium to the free flow is limited to the near-wall region, even for the highest porosity case. On the other hand, the causal interaction in the `top-down' direction is significantly stronger, and there is still observable causality between a higher layer of the channel flow (up to $y/H=0.2$) and flow motions at top layer pores.

\section{Acknowledgement}
The study has been financially supported by SimTech (EXC 2075/1–390740016) and SFB-1313 (Project No.327154368) from Deutsche Forschungsgemeinschaft (DFG). The authors gratefully appreciate the access to the high performance computing facility \emph{Hawk} at HLRS, Stuttgart of Germany.

\bibliography{reference}
\bibliographystyle{spbasic}

\end{document}